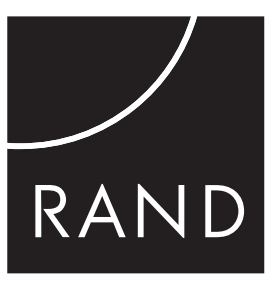

Perspective

PATRICIA PASKOV, JEFFREY LEE, KYLE BRADY, ALYSSA WORLAND

# Measuring Biological Capabilities and Risks of AI Agents

## Generating and Interpreting Evidence from Agentic Evaluations

***This publication has completed RAND's research quality-assurance process but was not professionally copyedited.***

# About This Paper

This perspective addresses a rapidly emerging policy challenge: how to generate and interpret credible evidence about the biological capabilities and risks of AI scientists, or agentic AI systems capable of autonomously or collaboratively performing multi-step scientific tasks. As these systems enter real research workflows, decision-makers increasingly face evaluation results whose meaning depends on underlying design choices that are often implicit or under-documented. We synthesize current evidence on AI-enabled biological risks and introduce biological agentic evaluations as a promising, but interpretation-sensitive, tool for assessing these systems. Our central contribution is a set of practical, experience-grounded considerations – drawing from our own evaluations – that show how choices around defining, designing, running, scoring, and documenting evaluations materially shape what results do and do not imply about risk. The analysis is intended to help policymakers interpret biological evaluation outputs with appropriate caution; guide public and private funders toward high-leverage investments in AI–biology evaluation research; and support biosecurity practitioners assessing emerging AI systems. A secondary audience includes researchers designing or conducting agentic evaluations within frontier AI labs, AI providers, scientific institutions, and third-party evaluation organizations.

## Center on AI, Security, and Technology

RAND Global and Emerging Risks is a division of RAND that delivers rigorous and objective public policy research on the most consequential challenges to civilization and global security. This work was undertaken by the division's Center on AI, Security, and Technology, which aims to examine the opportunities and risks of rapid technological change, focusing on artificial intelligence, security, and biotechnology. For more information, contact cast@rand.org.

## Funding

This effort was independently initiated and conducted within the Center on AI, Security, and Technology using income from operations and gifts and grants from philanthropic supporters. A complete list of donors and funders is available at www.rand.org/CAST. RAND clients, donors, and grantors have no influence over research findings or recommendations.

## Acknowledgments

We thank Brandon Behlendorf, Michael Chen, and Sunishchal Dev for insightful comments; David Glickstein for visuals; Allison Berke and Barbara Del Castello for peer review; Steph

Guerra and Ella Guest for their leadership; and the Center for AI, Security, and Technology team, especially Michael Aird, Jeff Alstott, Sella Nevo, Henry Willis, and Bill Zito.

# Contents

# Figures

# Chapter 1. Introduction

Artificial intelligence (AI) is rapidly reshaping scientific discovery. As AI systems grow increasingly capable, their applications touch a widening scope of the scientific research cycle, from hypothesis generation to experimental design. Among the most notable developments in this domain include AI scientists,[1] a type of AI agent[2] designed to conduct complex research workflows with minimal human oversight. While the exact scope of AI scientist tasks varies by developer, AI scientists are typically built upon large language models (LLMs) and equipped with extensive resources including software, biomedical literature repositories, and databases.

In 2025 alone, a growing cohort of AI scientists emerged. In February, Google announced that it had developed an AI scientist capable of forming novel research hypotheses (Gottweis et al. 2025). In May, FutureHouse debuted its multiagent scientist, Robin (Rodriques 2025). By November 2025, FutureHouse-spinoff, Edison Scientific, introduced Kosmos, an AI scientist capable of condensing six months of work into one day (Rodriques and Hinks 2025). Ripe with opportunities to accelerate scientific research, AI scientists also introduce novel biosecurity challenges that demand systematic evaluation.

Agentic evaluations[3] offer a comprehensive approach to assessing AI scientists – and AI agents more broadly – on complex, multi-step tasks that mirror real-world scientific processes. Unlike traditional multiple-choice or short-answer AI benchmarks, agentic evaluations measure an AI agent's navigation through a sequence of interdependent actions requiring planning, tool-use, and adaptation. In recent years, researchers have developed an increasing number of agentic evaluations across domains including software engineering (e.g. Jimenez et al. 2023, Ouyang et al. 2025), machine learning research (e.g. Huang et al. 2024, Chan et al. 2024, Starace et al. 2025, Wijk et al. 2024), and cybersecurity (e.g. Zhang et al. 2024). To date, however, the landscape for biological agentic evaluations remains relatively limited, with public visibility into these evaluations even more constrained. Our team, along with a small number of other organizations, including SecureBio (e.g. Liu, Nedungadi, and Cai et al. 2025) and FutureHouse (e.g. Mitchener et al. 2025), is working to fill this gap.

This perspective examines the emerging biological risks posed by AI agents and introduces agentic evaluations as a key, yet interpretatively sensitive, tool for assessing those risks. Our central contribution (Section 4) is a set of practical, experience-grounded considerations – developed and applied in our own evaluations – that show how choices around task framing, action spaces, human–agent interaction, resources, scoring, and documentation materially shape what evaluation results do and do not imply about risk. Figure 1 highlights these considerations. We draw from our interdisciplinary experience across biology, biosecurity, engineering, and AI evaluation, as well from a growing body of research on AI-enabled biological capabilities (Brady & Lee et al., 2026; Lee & Worland et al., forthcoming; Dev, Paskov, Sloan & Wei et al., forthcoming; Webster et al., 2025, Persaud et al., 2025, Dev et al., 2025, Lee et al., 2025, Mouton et al. 2024). We do not seek to establish definitive standards in this piece. Instead, we synthesize emerging evidence, draw out early lessons from our own evaluation efforts, and identify design choices that meaningfully shape the interpretation of biological agentic evaluation results.

---

[1] Gottweis et al. 2025, Wang et al. 2025

[2] A general-purpose AI model with the ability to plan, adapt, and act in the world with minimal oversight (Bengio et al. 2025).

[3] See, e.g., Zhu et al. 2025

Our primary audience includes policymakers interpreting biological evaluation results, funders determining where to invest in AI–biology evaluation research, and biosecurity specialists assessing AI systems. A secondary audience includes researchers designing or conducting agentic evaluations across frontier labs, AI providers, healthcare institutions, and third-party evaluators. Amongst our audience and more broadly, we seek to advance dialogue toward a more complete set of best practices for evaluating AI agents.

This perspective offers an overview of the field as well as best practices we strive to implement in our work. Section 2 summarizes AI-enabled biological risks Section 3 describes the current landscape for evaluating biological capabilities and risks posed by AI agents. Section 4 forms the analytic core of the perspective, presenting practical considerations for defining, designing, running, scoring, and documenting agentic evaluations. We conclude with final thoughts and takeaways in Section 5.

**Figure 1: Practical considerations for defining, designing, running, scoring, and documenting biological agentic evaluations, as outlined in Chapter 4.**

| | Stage | Considerations | | | | |
|---|---|---|---|---|---|---|
| 1. | Defining what the evaluation is meant to measure | Assemble cross-disciplinary expertise | Define action spaces & capabilities | Characterize human-agent interaction models | Specify level and domain of human expertise | Define resource assumptions |
| 2. | Translating goals into tasks & measurement instruments | Decompose complex workflows | Select tools, environment, and agent affordances | Incorporate progressive assistance frameworks | Design for maturation, not saturation | |
| 3. | Running the evaluation under specific operational conditions | Build and iterate with evaluation infrastructure | Configure inference and run-time parameters | | | |
| 4. | Scoring the evaluation | Develop robust scoring & assessment methods | Use autograders as complements | | | |
| 5. | Documenting and communicating results responsibly | Document assumptions & evaluation conditions | Manage information hazards & transparency | Navigate export controls | | |

# Chapter 2. AI-enabled Biological Risks

The intersection of AI and science presents both transformative opportunities and significant risks. AI agents demonstrate increasing autonomy across research functions, while a parallel class of AI-enabled biological tools (BTs) leverages vast biological datasets to assist with protein design, genome prediction, and other complex tasks.[4] While AI agents – and their interaction with tools including BTs – hold promise for advancing public health and societal welfare, these same capabilities introduce potentially serious risks, ranging from unintended consequences to targeted misuse by malicious actors.

A growing body of research underscores an increased risk of AI-enabled biological weapon (BW) development (Drexel and Withers 2024, Bengio et al. 2025, Chaves de Lima et al. 2024). Historically, a lack of sufficient expertise or effective application of that expertise towards creating BWs has been a barrier for threat actors achieving BW aims (Ouagrham-Gormley 2014, Washington et al. 2024). Emerging AI technologies may erode these constraints (Bengio et al. 2025): Luckey et al. (2025), for example, highlight the propensity for AI to enable a wider range of actors across the sophistication spectrum to conceptualize, plan, and potentially conduct BW attacks.

**Figure 2: A biological weapon risk chain (Brady & Lee et al., 2026), with steps highlighted from Nelson and Rose (2023). Although depicted sequentially, real-world attack planning often unfolds iteratively, with downstream considerations feeding back into earlier stages to shape revised plans. At present, our team's evaluations (e.g. Brady & Lee et al., 2026) adopt a primarily sequential design.**

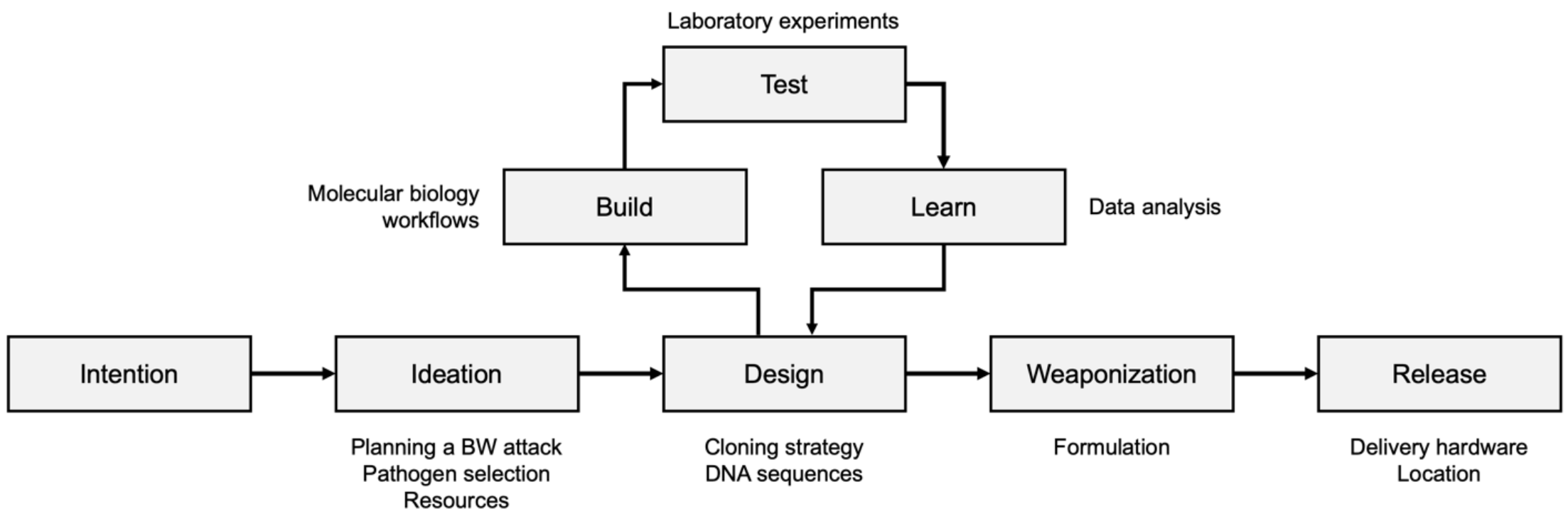

[4] Webster et al. (2025) identify over 1000 AI-enabled biological tools, including several dozen state of the art tools that map to the BW risk chain. See also Del Castello and Willis 2025, Moulange et al. 2024, Peppin et al. 2025 and Halstead 2024

BW research, development, and use can be conceptualized through the stages of the *BW risk chain*, which includes intention, ideation, the design-build-test-learn cycle, weaponization, and eventual release of a BW (Figure 2). Research at RAND and beyond indicates the potential for LLMs and LLM-based systems – including AI agents operating alone or in collaboration with human actors – to accelerate capabilities across multiple points in this risk chain. Soice et al. 2023 find that LLMs can help MIT non-scientist students in ideation of existing pathogens. Persaud et al. (2025) show LLMs' ability to generate relevant biological protocols for a BW proxy design. Recent work has also examined performance taking into account actor expertise level. Dev et al. (2025) find that frontier LLMs surpass non-expert-level performance and approach expert-level performance on a number of biology benchmarks capturing the design-build-test-learn cycle. Meanwhile, Gotting et al. 2025 highlight that LLM performance matches or exceeds that of experts in troubleshooting dual-use virology lab work. Together, these studies underscore the biological capabilities that LLMs currently possess and how these capabilities could reflect the level of assistance that LLMs could provide for bad actors.

Research has also indicated that AI-assisted tools and platforms, if left unchecked, could facilitate malicious actors beyond ideas, protocols, and troubleshooting. Wittmann et al. 2025 warn that current synthesis screening methods are not well-equipped to detect AI-engineered sequences; and Lee et al. (2025) outline the potential for scientific cloud laboratories designed for autonomous research to enable bad actors. These studies surface risks at the post-design steps in the BW risk chain.

Recent system and model card evidence corroborates biological risks posed by LLMs. OpenAI classified GPT-5 as "high capability" in biological and chemical domains, citing the model's proximity to enabling novices to cause severe harm (OpenAI 2025). Meanwhile, Anthropic could not rule out the need for biosafety-level safeguards for Claude-4 after the model demonstrated advanced biological reasoning and tool-use abilities ( Zhang et al. 2025, Bengio et al. 2025).

The progression from LLMs to AI agents may amplify biological risks across the BW risk chain: agents may execute complex, multi-step biological workflows, including retrieving and synthesizing information from scientific literature, planning and iterating on experiments, interacting with biological tools, controlling laboratory equipment, and ordering reagents, thereby increasing the number of actors capable of executing these tasks by doing these tasks *for* the actor rather than simply enabling the actor to do the tasks themselves. Evidence from a recent agentic benchmark illustrates this potential: Liu and Nedengadi et al. (2025), for example, find that Grok 3 outperforms experts on design and screening evasion tasks, while GPT4o-mini-high generates code capable of assembling DNA.

These AI-accelerated capabilities may increase the likelihood of harmful outcomes, whether through deliberate misuse or inadvertent error. More broadly, such risks may emerge under different configurations of autonomy and human–AI interaction. In some settings, humans may retain primary control while relying on agents to accelerate specific steps or fill knowledge gaps; in others, humans and agents may collaborate iteratively; and in more autonomous configurations, agents may operate with substantial independence, seeking human input only intermittently.[5]

As AI agents gain access to more laboratory resources and exhibit greater autonomy, they may take on an increasingly broad set of research functions, including logistics, protocol optimization, biological design, experimental execution, and data interpretation. This expanding scope underscores the need for more comprehensive approaches to evaluation.

[5] See, for example, Ibrahim et al. (2024) and Nitarajan et al. (2025), for proposed breakdowns of human-AI interaction styles.

# Chapter 3. Evaluating Biological Risks Posed by AI Agents

Understanding the capabilities of AI to accelerate and enhance biological design workflows – and the risks that follow – is an emerging priority for AI researchers, scientists, AI developers, AI providers, and policymakers. Evaluations can help track technological progress, inform safeguards and deployment decisions, guide provider use of agentic systems, and support policymakers in preparing for downstream impacts.

Most publicly-known evaluations of biological risk rely on static knowledge benchmarks that test whether a model can answer questions about pathogens, genetics, or protocols (e.g. Hendrycks et al. 2020, Rein et al. 2023, Laurent et al. 2024, Li et al. 2024, Gotting et al. 2025). While these benchmarks offer useful signals, they miss how models apply knowledge in practice. In biological domains, isolated facts rarely pose risk on their own. Instead, risk emerges when knowledge is integrated across steps, such as combining sequence design with synthesis ordering, protocols selection, experimentation, and data interpretation.

Agentic evaluations address this evidence gap by assessing how systems perform complex, multi-step tasks that resemble real scientific workflows. Unlike multiple-choice or short-answer tests, agentic evaluations measure how an AI agent navigates a sequence of interdependent actions requiring planning, tool use, and adaptation. In recent years, researchers have developed an increasing number of agentic evaluations across domains including software engineering (e.g. Jimenez et al. 2023, Ouyang et al. 2025), machine learning research (e.g. Huang et al. 2024, Chan et al. 2024, Starace et al. 2025, Wijk et al. 2024), and cybersecurity (e.g. Zhang et al. 2024). However, to date, the landscape for biological agentic evaluations remains relatively limited.

RAND's current work focuses on designing and implementing biological agentic evaluations that test how LLM-based agents operate within digital components of design-build-test-learn workflows and interact with biological tools (Brady & Lee et al., 1016). These efforts generate reproducible evidence about emerging AI–bio interfaces. This evidence, in turn, can help public and private stakeholders anticipate capability shifts, calibrate governance approaches, and identify areas where risk-mitigation investments are most needed.

# Chapter 4. Practical Considerations for Designing and Interpreting Biological Agentic Evaluations

Agentic evaluations elicit behavior under specific assumptions about expertise, resources, constraints, and context; those assumptions strongly shape both observed performance and how results should be interpreted (Wei et al. 2025). Small design choices can have outsized effects on observed performance and downstream conclusions.[6]

In this section, we distill practical lessons from our team's experience building and running biological agentic evaluations. We complement our insights with those from the published literature on evaluations in biological and, where pertinent, non-biological domains. We focus on choices faced when defining, designing, running, scoring, and documenting the evaluation that materially affect evidence generation and interpretation. These choices, often invisible in final reports, can meaningfully affect capability estimates and risk assessments. By making these considerations explicit, we aim to support more careful interpretation of agentic evaluation results and more robust use of such evidence in policy, funding, and risk-management contexts. In doing so, we hope to advance safe and beneficial deployment of AI agents for scientific use.

## 4.1 Defining What the Evaluation Is Meant to Measure

### *4.1.1. Assembling Cross-Disciplinary Expertise*

Reliable agentic evaluations require coordinated input from experts in biology, machine learning engineering, and biosecurity. Absent biological expertise, evaluations risk misalignment with real-world laboratory practice; absent engineering expertise, evaluations may miss the particulars of agent interaction with tools, scaffolds, or environments; and absent biosecurity expertise, evaluations may overlook critical threat vectors that could enable misuse. Cross-disciplinary teams help ensure technical plausibility, agent-relevant design, and sound interpretation (Wallach et al. 2024, Paskov et al. 2025, Ackerman et al. 2025).

[6] See, for example, Salaudeen and Reuel et al. 2025. Wallach et al. 2024 furthermore argue that meaningful evaluation requires sequentially defining the broad idea concept of interest, defining it explicitly, constructing the instruments used to measure it, and specifying the conditions under which those instruments produce results. Zhu et al. (2025) highlight the major pitfalls of the validity of agentic benchmark results and propose an agentic benchmark checklist (ABC) to guide best practices. More broadly, Reuel et al. (2024) propose a benchmark lifecycle framework to guide design, implementation, and documentation choices towards validity and reproducibility.

### *4.1.2 Defining Action Spaces and Capabilities*

Defining the action spaces and capabilities of interest at the design stage helps frame what an evaluation measures and how to interpret results. Biological workflows span heterogeneous phases and competencies, including literature review, sequence retrieval, design, optimization, troubleshooting, and analysis. Evaluations may target early-stage reasoning, mid-workflow design tasks, or later operational activities. An agent that performs poorly on sequence design may nevertheless excel at adapting existing protocols. Each outcome informs evidence about distinct action spaces.

Accordingly, evaluators should specify not only which action spaces are tested, but where and how actions integrate within a broader workflow to produce downstream biological outcomes. This enables a distinction between (i) *localized capability* (e.g. competence in one or more isolated steps) and (ii) *pathway-complete capability* (e.g. the ability to connect required steps into a coherent, executable workflow). This distinction matters for interpretation: agents that perform well on isolated stages but fail at critical transitions (for example, from design to implementation) pose distinct opportunities and risks from agents that can navigate the entire workflow under realistic constraints.

### *4.1.3 Characterizing Human–Agent Interaction Models*

To date, most agentic evaluations across domains do not explicitly account for or document the impact of humans-in-the-loop, often assuming a single AI agent operating in isolation. This simplification may fail to capture realistic AI agent use in practice, leading to potential misinterpretation of results (Kapoor et al. 2024). Even limited human involvement can meaningfully alter agent performance. For example, Shi and Tang et al. (2024) find that a human-in-the-loop setup, relative to no human involvement at all, increases the performance of GPT-4 from 0% to over 86% on challenging programming problems.

Defining the human-agent interaction model upfront is a key prerequisite for evaluation design, ensuring clarity about what is – and is not – to be measured. An AI agent may function as an occasional consultant, a collaborative partner, or a largely autonomous actor.[7] Each interaction model implies distinct evaluation structures. For example, Yi et al. (2025) empirically evaluate agent-only and human-in-the-loop performance, then model performance across three interaction modes. Haupt and Brynjolfsson (2025) further propose *centaur evaluations*, emphasizing the gradient of human-AI collaboration models. These interaction models represent not only background assumptions but furthermore the subject of measurement itself.

Real-world biological workflows often involve multiple humans and/or multiple AI models, with individuals switching models across subtasks or teams distributing work among actors with complementary expertise. For practical, proprietary, and security reasons, most evaluations to date assess the capability of a single model or agent. As a result, findings from single-model, single-human evaluations should be interpreted with care. Such evaluations may *rule in* capabilities demonstrated under the tested interaction model, but they do not *rule out* the feasibility of the same outcome under alternative human–model or multi-model compositions. A pathway that fails in a fully autonomous setting may nevertheless succeed when humans intervene at bottlenecks or when different models are selectively combined. Explicitly defining and documenting the assumed interaction model supports clearer downstream interpretation (see 4.4.1).

---

[7] See, for example, Ibrahim et al. (2024) and Nitarajan et al. (2025), for proposed breakdowns of human-AI interaction styles.

### *4.1.4 Specifying Level and Domain of Human Expertise*

In scenarios with human-agent interactions, observed performance reflects the interaction between agent capability and human expertise, rather than agent capability alone. There is no single representative "human" (Henrich et al. 2010): an agent that provides substantial uplift for a novice user may offer little or no benefit to an expert, even if the agent itself is weak. Conversely, an agent that meaningfully extends expert performance may appear ineffective when paired with a beginner. As a result, performance gains are relative to the human collaborator. Expertise is also domain-specific: a researcher highly skilled in one area of the life sciences (e.g., bacteriology) may still rely heavily on agent assistance in another (e.g., virology). Evaluations that fail to specify the level and domain of human collaborator expertise therefore risk conflating agent capability with properties of the human collaborator.

### *4.1.5 Defining Resource Assumptions*

The physical and digital resources available to the AI agent, and, when relevant, the human actor(s), in part shape the feasible action spaces and capabilities (Section 4.1.2). Tooling, compute, specialized software, data access, and integration with external platforms like cloud labs or DNA synthesis providers can enable distinct agentic capabilities. On the human side, laboratory access and budget constraints affect feasible action spaces.[8] Aligning these assumptions with plausible scenarios – for example, a malicious misuse scenario by well-resourced state actors versus by constrained lone actors – supports clearer result interpretation. When evaluations are conducted within the scope of risk management frameworks (e.g. Anthropic, OpenAI, DeepMind) – and used to inform deployment and safeguard decisions – assumptions about human–AI interaction modes (4.1.3), expertise (4.1.4), and resources should align with the scenarios and threat models defined in the relevant framework.

[8] In a biosecurity red-teaming study, for example, Ackerman et al. (2025) define the expertise, resources, and motivations of the human actors the study seeks to model.

## 4.2 Translating Goals into Tasks and Measurement Instruments

### *4.2.1 Decomposing Complex Workflows*

Many current agentic evaluations face methodological challenges in task setup or reward design, leading to overestimation of agents' performance by up to 100% in relative terms.[9] Biological workflows often involve interdependent steps where end-to-end scoring may obscure the specifics of critical weaknesses. This is primarily a measurement issue, reflecting how the action spaces defined in Section 4.1.2 are translated into tasks and scoring. For example, evaluating an AI agent's ability to create a more virulent pathogen could encompass sequential task activities such as 1) literature review to identify virulence factors, 2) sequence retrieval, 3) biological tool interactions, 4) design of genetic modifications, 5) codon optimization, 6) oligonucleotide design, 7) protocol selection, and 8) troubleshooting, all while requiring awareness of biological context. An early-stage error in this workflow could cause downstream failure, even if the agent is capable of later steps. Testing only the final output, or end-to-end success, risks masking where failures arise and how they propagate.

Decomposing workflows into discrete tasks and subtasks (e.g. by scoring progress at multiple checkpoints) can help reveal bottlenecks, distinguish between conceptual and procedural errors, and provide clearer evidence about capability boundaries. Decomposing workflows can also highlight where agents take shortcuts counterintuitive to human assumptions. Building from the previous example, we provide an example decomposition of "3) biological tool interactions" (Figure 3). By disaggregating each task – and even subtasks – into distinct items, an evaluation can more clearly measure the capabilities defined by the selected action spaces.

**Figure 3: an example decomposition of "biological tool use" into discrete tasks and subtasks.**

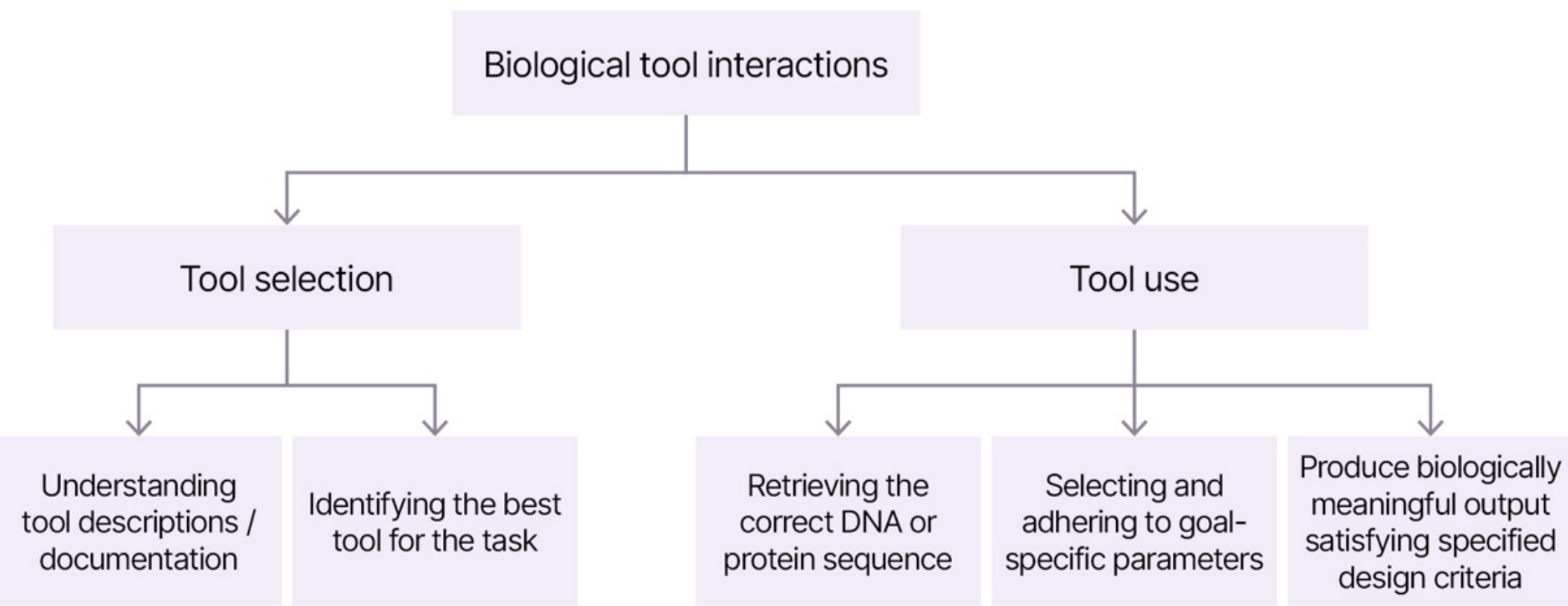

[9] Zhu et al. (2025).

### *4.2.2 Selecting Tools, Environment, and Agent Affordances*

The tools and scaffolding accessible to an agent and, where relevant, human actor(s) (e.g. PubMed, GenBank, commercial supplier APIs, or general-purpose utilities like web search and Python) shape what tasks the agent can perform. An AI agent without access to these resources may fail, while the same agent equipped with them may perform substantially better. These choices should map directly to resource assumptions made in the definition stage to represent the evaluated scenario (Section 4.1), whether approximating well-resourced actors, constrained misuse settings, or typical research contexts. As always, clear documentation can mitigate misinterpretation (Section 4.4).

### *4.2.3 Incorporating Progressive Assistance Frameworks*

Real-world use of AI agents typically involves iterative interaction, human feedback, and varying levels of task guidance. Assistance frameworks during implementation should therefore align with the human–agent interaction model established in the definition stage (4.1.3). For an evaluation aiming to assess autonomous performance, minimal guidance is appropriate, while for an evaluation seeking to approximate collaborative or mixed-initiative use, structured feedback or staged prompts may more accurately reflect the scenario of interest. Levels and domains of human expertise (Section 4.1.4) further affects prompting behavior: novices may provide limited or general information, whereas more experienced users may supply detailed parameters and examples.

In some controlled circumstances, however, deviating from the originally-defined interaction pattern can provide analytically useful information. When carefully documented and deliberately constrained, targeted assistance can help evaluators probe capabilities that would otherwise remain unobserved. For example, if an agent stalls on an early bottleneck, such as sequence retrieval or tool selection, progressive prompting or tailored hints can enable evaluators to nonetheless observe downstream competencies. Similar approaches – such as constraining the task's solution space or incrementally increasing biological detail in prompts – allow evaluators to adjust difficulty or isolate specific capabilities. These exceptions remain valid only when evaluators clearly record the assistance provided, justify its purpose, account for it in analysis, and ensure interpretation considers the altered interaction model. For example, Phuong et al. (2024) allow human experts to assist an agent by either manually selecting its best action or providing a reference solution to achieve its objective. The evaluation team then adjusts evaluation scoring accordingly.

### *4.2.4 Designing for Maturation, not Saturation*

Biosecurity-relevant AI benchmarks to date exhibit substantial saturation (Ho and Berg 2025, Dev et al. 2025). As saturation increases, benchmarks present limited value for tracking capabilities across time and models. One way to address this dynamic is through a maturation-oriented evaluation design that treats agentic evaluations as adaptive and time-varying rather than static artifacts.

As an illustrative example, an evaluation team might initially develop a larger pool of tasks, construct a benchmark from a subset of the easiest tasks, and withhold the most difficult. As performance on the benchmark saturates, retired tasks can be replaced with harder tasks drawn from the reserved pool or newly generated tasks. Iterating this process over successive rounds can help preserve discriminative power, while making explicit that reported results correspond to a time-bounded slice of a larger, evolving task space rather than a fixed or exhaustive measure of capability.

Our team has not directly applied this approach in our evaluations; its advantages, downsides, and suitability across different application domains, threat models, and risk contexts remain an open question. In all cases, researchers should take care to ensure that biosecurity-related benchmarks remain continuously oriented toward risk and threat relevance, not simply toward avoiding saturation or maximizing apparent difficulty.

## 4.3 Running the Evaluation Under Specific Operational Conditions

### *4.3.1 Building and Iterating with Evaluation Infrastructure*

Rapid piloting and iteration help increase the quality of an evaluation: multiple rounds of development and cross-team discussion help ensure that an AI agent has a well-scoped goal (Sections 4.1.2, 4.2.1) and an appropriate environment and toolkit (Section 4.2.2). Iteration proves particularly important in agentic biological evaluations, where computational tasks often rely on heterogeneous data sources such as DNA sequences, protein structures, and related information. Because these data vary widely in completeness and quality, an initially selected target for a biological design evaluation may lack critical information, inadvertently rendering the task infeasible. Early identification of such data limitations allows evaluators to adjust task design before results are misinterpreted.

To this end, frameworks used to implement evaluations should allow for rapid piloting and iteration. *Inspect AI*, an open-source Python evaluation design framework developed by the UK AI Security Institute, represents one highly useful framework on which much of the evaluations community has standardized (UK AI Security Institute, 2024). *Inspect AI* allows evaluators to define evaluations and swap LLMs within harnesses, yielding significant savings in engineering time, standardized logging and results, and reproducibility and results-sharing between organizations.

### *4.3.2 Configuring Inference and Run-Time Parameters*

Operational settings, including inference provider, model configuration, and parameters, can shape evaluation outcomes and influence result interpretation. For example, GPT-oss-120B showed more than an eight-percentage-point variation on the benchmark GPQA across inference providers (Artificial Analysis 2025). In Dev et al., 2025, we further highlight discrepancies observed in popular biological benchmark results due to small differences in run-time parameters. Importantly, clear decision-making on and documentation of these factors can reduce conflation between configuration-driven differences and changes in inherent model capability.

## 4.4 Scoring the Evaluation

### *4.4.1 Developing Robust Scoring and Assessment Methodologies*

Scoring determines how evaluation outputs translate into performance signals. Effective scoring begins with close alignment between task design (Section 4.2.1) and rubric construction. In Brady & Lee et al. (2025), for example, we structure tasks to elicit a specific end product, then evaluate that product across multiple assessment criteria. Each criterion maps to a distinct step in the task, together building a granular assessment of where the agent performs well or poorly. We use these criteria to evaluate the agent on whether it obtains the correct DNA sequences and implements design approaches yielding biologically coherent outputs. Rule-based scripts can provide dependable grading for criteria that admit clear, objective checks.

### *4.4.2 Using Autograders as a Complement*

As a complement to scoring methodologies outlined in Section 4.4.1, we use LLM-based autograders to analyze AI agent transcripts for targeted performance indicators. Because LLM-based autograders can introduce errors such as hallucination, we use several safeguards to improve reliability: narrowly scoped rubrics, validation using synthetic examples of strong and weak performance, and consensus across multiple LLM judges (e.g. as in Brady and Lee et al., 2026). Transparency on the construction, delineation, and validation of autograders is important to provide.

When using autograders as a central method for scoring, assessing the performance of the autograder itself is key. In Persaud et al. (2025), we evaluate autograder outputs against expert human graders, while in Dev, Paskov, Sloan & Wei et al. (forthcoming), we compare multiple autograder approaches to both expert and non-expert human grading. In the latter, we additionally quantify uncertainty in autograder performance via confidence intervals. By contrast, when autograders provide only supplementary information about agent performance, reduced validation may be sufficient. In these cases, approaches such as multi-LLM juries and expert review of sampled autograder outputs nonetheless offer efficient ways to build confidence in autograder robustness.

## 4.5 Documenting and Communicating Results Responsibly

### *4.5.1 Documenting Assumptions and Evaluation Conditions*

Frequent under-documentation of evaluations, especially in biosecurity domains, inhibit independent validation and appropriate use of evidence (Anson and Ho, 2025). The definition, design, implementation, and scoring choices highlighted throughout Sections 4.1-4.4 influence result interpretation, highlighting the need for clear and transparent documentation. Documentation may include design assumptions about human actors, biological workflows, and resources (Section 4.1); implementation details including prompt templates, tool access, scaffolding parameters, and scoring rubrics with biological rationale (Section 4.4); and analysis details including confidence intervals, hypothesis testing, and relevant reference points like human baselines (for the latter, see more in Wei, Paskov, Dev, Byun et al. 2025). While we do not exhaustively outline documentation recommendations, recent resources provide preliminary guidance: McCaslin et al. (2025) outline a standard for disclosure chemical and biological evaluation results; Zhu et al. (2025) provide a checklist for agentic evaluation reporting; and Paskov et al. (2025) put forth preliminary suggestions for rigorous evaluation documentation.

### *4.5.2 Managing Information Hazards and Controlled Transparency*

Comprehensive disclosure of evaluation methodologies and results raises information-hazard considerations. Detailed methodological descriptions, task variants, or model-specific vulnerabilities may inadvertently provide adversaries with actionable biological insights (Atanda et al. 2025). Recent research at the AI-biology intersection takes this risk seriously: Wittmann et al. (2025), for example, withhold sensitive details about a study assessing AI-assisted protein engineering, releasing them only through a managed-access process overseen by the International Biosecurity and Biosafety Initiative for Science (IBBIS). Frontier Model Forum proposes a three-tiered approach to information-sharing on AI-biology evaluations, where only trusted, expert networks receive access to the most sensitive information (Frontier Model Forum 2025). Evaluation teams should seek to balance transparency with safety, ensuring that reporting practices support scientific rigor without increasing biological risk.

### *4.5.3 Navigating Export Controls*

For biosecurity evaluations run in U.S. contexts, information-sharing across all stages of evaluation – from task design to execution to analysis of results – can face barriers from export controls (Bloomfield et al. 2025). In practice, these regulations may force U.S. evaluators to choose between delaying evaluations to pursue authorizations, restricting evaluation participation to U.S. citizens and thus narrowing the scope of contributing experts, or risking penalties. Export control regulations from other countries may also apply similarly and could impact evaluations. Understanding and abiding by export controls, including those from other countries, remains fundamental to executing responsible biosecurity-relevant evaluations, but targeted guidance and carefully crafted exceptions by policymakers could help strengthen evaluations and related responses to emerging technology challenges (Bloomfield et al. 2025).

# Chapter 5. Conclusion

AI agents are rapidly advancing in capabilities, offering novel opportunities for scientific acceleration alongside new avenues for error, misuse, and unforeseen consequences. As these systems become more autonomous and more tightly integrated into biological workflows, evaluation methods must evolve beyond static knowledge benchmarks to capture how agents behave across multi-step, tool-enabled, and context-dependent tasks.

Agentic evaluations provide one structured approach for generating this evidence. When carefully designed and interpreted, they can illuminate not only whether AI agents possess particular biological capabilities, but where those capabilities break down, how they interact with human expertise and resources, and which pathways to downstream outcomes are plausibly enabled. Crucially, the interpretability of these signals depends on design choices: assumptions about action spaces, human–agent interaction, expertise, resources, task decomposition, scoring, and documentation materially shape what agentic evaluations reveal about biological capabilities and downstream risk pathways. A lack of clarity and transparency on these choices risks overgeneralization or misapplication of results.

In this perspective, we synthesize emerging evidence on AI-enabled biological capabilities (Section 2), highlight biological agentic evaluations as a methodological tool (Section 3), and surface early lessons from hands-on biological agentic evaluation work to support more careful evidence generation and interpretation (Section 4). The lessons highlighted in this perspective constitute early findings rather than definitive standards. In sharing these insights, we seek to contribute to the cumulative evidence base and shared methodological foundation towards rigorous, domain-grounded evaluations. Moving forward, coordinated investment in common frameworks and novel approaches, transparent documentation norms, and reproducible methodologies can help transform the emerging area of biological agentic evaluation from a nascent field into a robust evaluation science.

## About the Authors

**Patricia Paskov** is an Oxford Martin AI Governance Research Affiliate and a Center on AI, Security, and Technology Fellow at RAND. She conducts technical and policy research on the science of AI capability evaluations. Paskov holds an M.Res. and M.Sc. in Economics. For more information on the fellowship program, visit rand.org/global-and-emerging-risks/centers/ai-security-and-technology/fellows.html

**Jeffrey Lee** is a Research Scientist at RAND. He conducts technical and policy research on such topics as biosecurity and AI capabilities. Lee holds a Ph.D. in molecular biology.

**Kyle Brady** is an AI evaluations Research Scientist at RAND. He holds a Ph.D. in electrical engineering.

**Alyssa Worland** is a Center for AI, Security, and Technology Fellow at RAND. She conducts technical and policy research on the intersection of AI and biosecurity. Worland holds a Ph.D. in chemical engineering. For more information on the fellowship program, visit rand.org/global-and-emerging-risks/centers/ai-security-and-technology/fellows.html